\documentclass[aps,prd,nofootinbib,superscriptaddress,twocolumn]{revtex4-1}

\usepackage[bookmarks=true]{hyperref} 
\usepackage{amsmath,amsfonts,amsmath}
\pdfoutput=1  
\usepackage{graphicx,subfigure,epsfig,color}

\newcommand{\ba}{\begin{eqnarray}}
\newcommand{\ea}{\end{eqnarray}}
\newcommand{\nn}{\nonumber}

\def \be {\begin{equation}}
\def \ee {\end{equation}}
\def \barr {\begin{array}}
\def \earr {\end{array}}
\def \bea {\begin{eqnarray}}
\def \eea {\end{eqnarray}}

\def \ble {\begin{widetext}\begin{equation}}
\def \ele {\end{equation}\end{widetext}}
\def \blea {\begin{widetext}\begin{eqnarray}}
\def \elea {\end{eqnarray}\end{widetext}}

\def \nn {\nonumber}

\newcommand{\eq}[1]{(\ref{#1})}

\def \e {\epsilon}

\def \Tr {{\textrm{Tr}}}

\def \and {{\textrm{and}}}

\begin{document}

\title{Compact Star of Holographic Nuclear Matter and GW170817}

\author{Kilar Zhang}
\email{kilar.zhang@gmail.com}
\affiliation{Department of Physics, National Taiwan Normal University, Taipei 11677, Taiwan}

\author{Takayuki Hirayama}
\email{hirayama@isc.chubu.ac.jp}
\affiliation{College of Engineering, Chubu University,
1200 Matsumoto, Kasugai, Aichi, 487-8501, Japan}

\author{Ling-Wei Luo}
\email{lwluo@gate.sinica.edu.tw}
\affiliation{Institute of Physics, Academia Sinica, Taipei 11529, Taiwan}

\author{Feng-Li Lin}
\email{fengli.lin@gmail.com}
\affiliation{Department of Physics, National Taiwan Normal University, Taipei 11677, Taiwan}

\begin{abstract}
 We use a holographic model of quantum chromodynamics  to extract the equation of state (EoS) for the cold nuclear matter of moderate baryon density.  This model is based on the Sakai-Sugimoto model in the deconfined Witten's geometry with the additional point-like D4-brane instanton configuration as the holographic baryons. Our EoS takes the following doubly-polytropic form: 
$ \epsilon=2.629 {\cal A}^{-0.192} p^{1.192}+0.131 {\cal A}^{0.544} p^{0.456}$
with $\cal A$ a tunable parameter of order $10^{-1}$, where $\epsilon$ and $p$ are the energy density and pressure, respectively. The sound speed satisfies the causality constraint and breaks the sound barrier.  We solve the Tolman-Oppenheimer-Volkoff  equations  for the compact stars and  obtain the reasonable compactness for the proper choices of $\cal A$. Based on these configurations we further calculate the tidal deformability of the single and binary stars. We find our results agree with the inferred values of LIGO/Virgo data analysis for GW170817.
\end{abstract}

\maketitle


\section{Introduction}

  Tremendous gravity can transform the ordinary matter in a compact star into exotic nuclear matter such as neutron liquid or quark-gluon plasma, which are hard to produce on earth and whose properties remain to be clarified after decades of studies  \cite{Lattimer:2012nd,Baym:2019yyo}. By the same token, the gravitational tidal force acting on the nuclear matter of a compact star can cause shape deformation, which can reveal nuclear matter's hydrodynamical properties such as equation of state (EoS). A novel way of observing the tidal deformation is to detect the gravitational wave emitted during the binary merger of compact stars such as neutron stars (NS) \cite{Rezzolla:2016nxn}. A recent example is LIGO/Virgo's GW170817 on the observation of gravitational wave from the merger of binary neutron stars (BNS)\cite{TheLIGOScientific:2017qsa,Most:2018hfd,Abbott:2018wiz,Abbott:2018exr}, which yields the masses of the neutron stars and the upper bound on the tidal deformability, and has inspired closer examination of the EoS, see for examples \cite{Tews:2018chv,De:2018uhw,Zhao:2018nyf,Han:2018mtj,Carson:2018xri,Tews:2019cap}.  One shall expect to observe more BNS events in the coming future to infer the more precise relation between mass and tidal deformability from parameter estimation, and reveal the candidate EoS.

   On the other hand, it is notoriously difficult to derive the EoS of exotic nuclear matters at moderate baryon density, i.e., about a few of the saturation density of nuclei,  either from the first principle such as lattice quantum chromodynamics (QCD) by suffering a sign problem at finite chemical potential \cite{deForcrand:2010ys,Cristoforetti:2012su}, or from perturbative  QCD and chiral perturbation theory due to a sizable coupling at moderate densities \cite{Kraemmer:2003gd,Kurkela:2016was,Vuorinen:2018qzx}. Therefore, most of EoS currently used for nuclear matter in compact stars are phenomenological. It is important to  derive EoS systematically based on some physical principle, and the holographic QCD is suitable for such a purpose.  
 
    Holographic QCD is an effective theory for QCD in terms of the dual bulk gravity dynamics based on the spirit of AdS/CFT correspondence \cite{Maldacena:1997re}. It has been adopted to address many QCD problems with success, e.g.,  clarify the hydrodynamical nature of quark-gluon plasma \cite{Kovtun:2004de,Liu:2006ug,Gubser:2006bz,Gynther:2010ed}  in the experiments of heavy ion collisions. Among many holographic QCD models, the Sakai-Sugimoto (SS) model  \cite{Sakai:2004cn,Sakai:2005yt}, where the mesons are introduced as the $D8/\overline{D8}$-branes (or called meson-branes) in the background of Witten's geometry \cite{Witten:1998zw}, is the best model so far  with very few free parameters.   
Especially, the SS model realizes the quark confinement and chiral symmetry breaking in a natural and geometric manner, and yields the chiral Lagrangian with well-fitted meson and  hadron spectra, and the decay amplitudes \cite{Sakai:2004cn,Sakai:2005yt,Hata:2007mb}.
 
    Our goal in this paper is to extract EoS of holographic nuclear matters from SS model, and use it to study the properties of the associated compact stars.  We start with the SS model in the deconfined Witten's geometry, in which both the broken and unbroken phases of chiral symmetry can be realized. This is suitable for the consideration of QCD at finite baryon density because the chiral symmetry is expected to be restored at high enough baryon density. The baryons are introduced as the D4-brane instanton \cite{Witten:1998xy,Hata:2007mb}. Here we will only consider the point-like instanton configuration \cite{Bergman:2007wp} which should be good enough approximation for the case of moderate baryon density.  Our EoS has only one tunable parameter, and by the proper choice we find the mass, radius and tidal deformability of the compact stars are in excellent agreement with the inferred values from the data analysis of GW170817.   
     
 This paper is organized as follows. In section II we briefly review the construction of  holographic nuclear matters as the point-like D4-brane instantons based on SS model in the deconfined phase. In section III we extract the EoS of nuclear matters at modest baryon density by numerically solving the thermodynamics of the holographic model constructed in section II. In section IV we then apply this EoS to solve for the compact stars and show the configurations are comparable with neutron stars. Moreover, we further compare the tidal Love numbers with the ones extracted from GW170817 observation and find excellent agreement. Section V is the conclusion.   
 
\section{Holographic nuclear matters}
   
   The meson-brane action of the SS model consists of the non-Abelian Dirac-Born-Infeld (DBI) action and the Chern-Simons (CS) action of the $D8$ and $\overline{D8}$-branes, \be\label{Sdbi}
 S_{DBI} =
\frac{2 T_8 \Omega_4 V_3} {T}  \int_{U_c}^\infty \! dU e^{-\Phi} \sqrt{ \det ( g +2\pi \alpha' {\cal F} ) },
\ee
and
\be\label{Scs}
S_{CS} =T_8\int C_3\textrm{Tr}(2\pi \alpha' {\cal F})^3=
 \frac{N_c V_3}{4\pi^2 T} \int_{U_c}^\infty \hat{A}_0\Tr[ F\wedge F],
\ee
where $g$ is the induced metric, $\alpha'$ is the string tension, $T_8$ is the tension of $D8$-brane, $\Phi$ is the dilaton, $\Omega_4$ is the volume of the unit 4-sphere on which the $D8$-branes wrap. The remaining worldvolume directions of $D8$ are the Euclidean 4-plane of temperature $T$ and volume $V_3$ where the dual QCD lives with assuming translational invariance, and the holographic coordinate $U$ ranging from the holographic boundary $U=\infty$ to the tip $U=U_c$ at which the $D8$ and $\overline{D8}$ connect.  The $N_c$ is the number of the background RR-flux $F_4=dC_3$ and is dual to the number of colors of QCD.  We will consider two-flavors QCD which is dual to a bulk U(2) gauge theory so that its field strength can be decomposed into a $U(1)$ and an $SU(2)$ part,
\be
{\cal F}_{\mu\nu} =\hat{F}_{\mu\nu}+F_{mu\nu}, \qquad F_{\mu\nu}=F^a_{\mu\nu}\sigma_a
\ee
where $\mu,\nu=0,1,2,3,U$ and $\sigma_a$ with $a=1,2,3$ the Pauli matrices, and the nonvanishing components are $\hat{F}_{0U}=-\partial_U \hat{A}_0$, $F_{ij}$ and $F_{iU}$ with the boundary value of $\hat{A}_0$ dual to the baryon chemical potential.
   
   For our study of holographic uniform nuclear matter, we will consider the SS model in the deconfined phase for which the background is the deconfined Witten's geometry, and the detailed form of its induced metric $g$ on the $D8$-brane can be found in \cite{Witten:1998zw,Bergman:2007wp,Li:2015uea}. This metric describes a black brane with a blacken factor
\be   
f_T=1-U_T^3/U^3 \nn
\ee
which corresponds to a Hawking temperature  $T=3U_T^{1/2}/4\pi R^{3/2}$ with $R$ the curvature radius of the Witten's geometry. The SS model in the deconfined phase is the holographic version of the NJL model \cite{Antonyan:2006vw,Davis:2007ka} which can be adopted for describing the phase of baryon fluid \footnote{The black brane is not the thermodynamically dominant phase over the confining cigar geometry at very low temperaure. Despite that, we will still extrapolate our results all the way to $T=0$ as currently there is no known holographic model for the baryon fluid in a confining geometry.}.

   The baryons in SS model can be introduced via Witten's mechanism \cite{Witten:1998xy}, which corresponds to adding the wrapped $D4$-brane instanton configuration on the $D8$ worldvolume. There are many different instanton configurations, for the most recent study see \cite{BitaghsirFadafan:2018uzs}. In this paper we will consider the simplest one, namely, the point-like instanton \cite{Bergman:2007wp} located at the tip $U=U_c$, \i.e., $\textrm{Tr}[F\wedge F] \propto N_I \delta(U-U_c)$, so that we also need to add a wrapped $D4$ action
\be\label{Sd4}
S_{D4}={N_I T_4 \Omega_4\over T} \int dU e^{-\Phi} \sqrt{g} \; \delta(U-U_c)
\ee 
where $N_I$ and $T_4$ are the number and the tension of instantonic $D4$-branes,  respectively.

   Denote the $D8$ profile by $X_4(U)$ where $X_4$ is the compact direction perpendicular to  $D8$, then plug it into \eq{Sdbi} and combine with \eq{Scs} and \eq{Sd4} to get the total action
\be\label{totalS}
S={{\cal N} V_3 \over T} \int_{u_c}^{\infty} {\cal L}, \qquad {\cal N}={N_c \lambda_{YM}^3 \over 12 (2\pi)^5} M_{KK}^4
\ee
with $\lambda_{YM}$ the 't Hooft coupling and $M_{KK}$ the mass gap of the holographic QCD \footnote{For the detailed relations between bulk quantities such as the string length, string tension, etc and the QCD quantities such as $\lambda_{YM}$, $M_{KK}$, etc, please see \cite{Sakai:2004cn,Sakai:2005yt}. With these relations, one can arrive \eq{totalS}.}, and 
\be\label{calL}
{\cal L}=u^{5/2}\sqrt{1-a_0'^2+u^3 f_T x_4'^2}+n_I [{u\over 3}\sqrt{f_T}-\hat{a}_0]\; \delta(u-u_c)
\ee
where $'=\partial_u$. In the above we have adopted the scaled convention introduced in \cite{Li:2015uea} so that $\cal L$ is expressed in terms of dimensionless lower-case variables such as $u$, $n_I$ and $\hat{A}_0$ instead of their dimensionful upper-case counterparts. 

  We can solve the equations of motion derived from \eq{calL} for $x_4$ and $\hat{a}_0$. Up to the first integration and requiring $x_4'\sim k u^{-11/2} + \cdots$ and $\hat{a}_0=n_I u^{-5/2}+\cdots$ at $u=\infty$ from the dictionary of AdS/CFT, the result is
\be\label{EoM}
 \hat{a}_0'=n_Iu^{3/2}\sqrt{f_T g^{-1}}, \quad x_4'=k u^{-3/2}\sqrt{f_T^{-1} g^{-1}}
\ee
with $k$ the integration constant and 
\be
g=(u^8+u^3 n_I^2) f_T-k^2. \nn
\ee
To further integrate  \eq{EoM}, it needs to impose the following boundary conditions
\be\label{BC2}
\mu=\hat{a}_0(\infty)=\hat{a}_0(u_c)+\int_{u_c}^{\infty} du\; \hat{a}_0', \quad \ell=2\int_{u_c}^{\infty} du\; x_4'
\ee
where $\mu$ is the baryon chemical potential and  $\ell$ is the (scaled) separation of $D8$ and $\overline{D8}$ at $u=\infty$.

  The solutions of \eq{EoM} and \eq{BC2} can be used to obtain the on-shell action, which yields the (dimensionless) dual grand canonical potential density \footnote{$\Omega_{\ell}$ is UV divergent and a regularization scheme is needed. We simply introduce an UV cutoff $u=\Lambda$ and subtract the $2\Lambda^{7/2}/7$.}
\be\label{Grandpotential}
\Omega_{\ell}[T, \mu; n_I,u_c]={1\over \cal N} {T\over V_3} S|_{\textrm{on-shell}}=\int_{u_c}^{\infty} du\; u^{13/2} \sqrt{f_T g^{-1}}\;.
\ee
Here $n_I$ and $u_c$ are not the thermodynamical variables but the parameters for the instanton configurations, with respect to which we need to minimize $\Omega_{\ell}$ to obtain some conditions to fix their values. These conditions ${\partial \Omega_{\ell}\over \partial n_I}={\partial \Omega_{\ell}\over \partial u_c}=0$ yield respectively
\be\label{Minimize}
\hat{a}_0(u_c)=u_c\sqrt{f_T(u_c)}/3, \quad n_I u_c^{3/2} (1+f_T(u_c))=6\sqrt{g(u_c)}.
\ee
Thus, by solving \eq{BC2} and \eq{Minimize} we can fix the values of $u_c$, $n_I$ and $k$ for a given set of $\mu$ and $T$ with a fixed model parameter $\ell$, and then determine $\Omega_{\ell}$.   

  Once the grand canonical potential $\Omega_{\ell}[\mu, T]$ is determined, through the thermodynamic relations we can determine the pressure $p_{QCD}$, baryon number density $n_I$ and  energy density $\epsilon_{QCD}$ as follows
\be\label{thermo}
p_{QCD} =-\Omega_{\ell},  \, \, n_I=-{\partial \Omega_{\ell} \over \partial \mu}|_T, \, \, \epsilon_{QCD} =\Omega_{\ell}+n_I \mu-T{\partial \Omega_{\ell}\over \partial T}|_{\mu}.
\ee
Thus, in principle the EoS can be obtained \footnote{We also need to make sure the phase of nuclear matter is dominated over the mesonic phase, i.e., $n_I=0$, and quark phase, i.e., with mesonic branes ending on horizon by comparing their free energies. }. 

 All the quantities in \eq{thermo} are dimensionless, and their dimensionful counterparts are given by 
\be\label{nEoS}
p=c^2 {\cal N} \ell^{-7} p_{QCD}, \quad n_b= {\cal N}^{3/4} \ell^{-5} n_I, \quad \epsilon={\cal N}\ell^{-7} \epsilon_{QCD}.
\ee
Here $c$ is the light speed. 

  We can see that $\ell$ and $\cal N$ are the only tunable parameters of our model. For convenience, we choose the typical values given in \cite{Sakai:2004cn,Sakai:2005yt} 
\be\label{calN}
\lambda_{YM} N_c \simeq 24.9, \; M_{KK} \simeq 949 \textrm{ MeV} \Rightarrow {\cal N}=1.2 \times 10^{10}  \textrm{ MeV}^4.
\ee
by which it yields well-fitted QCD spectra. However, from \eq{nEoS} we see that both $p$ and $\epsilon$ depend on only the combined form ${\cal N} \ell^{-7}$. Thus, different choice of $\cal N$'s value is just re-parametrizing the model parameter $\ell$ as far as only EoS is concerned.

\section{Equation of state}

   Based on the setup in the previous section, we can numerically obtain $p_{QCD}(\mu,T)$ and $\epsilon_{QCD}(\mu,T)$ and then extract the EoS by combining the result. We can fit the EoS into the piecewise polytropic form. Especially, we find that for the small value of $p_{QCD}\in [0,0.05]$, the EoS is well-fitted by the following doubly polytropic function (for $T=0$)\footnote{For comparison, the EoS for $T=0.05 M_{KK}$, i.e., $T=47.45 \textrm{ Mev}$ for $M_{KK}=949  \textrm{ Mev}$ is well-fitted by 
\be 
\epsilon_{QCD}=2.617 \; p_{QCD}^{1.188}+0.128 \; p_{QCD}^{0.452}
\ee
Note that this temperature is considerably high from typical astrophysical point of view for a neutron star, however, we see that its effect to EoS is quite small. Thus, we will neglect the temperature effect in the following discussions.}   
\be\label{EoSmain}
\epsilon_{QCD}=2.629 \; p_{QCD}^{1.192}+0.131 \; p_{QCD}^{0.456}, \quad p_{QCD}\in [0,0.05].
\ee  

\bigskip

   We will argue that $p_{QCD}\simeq 0.05$ corresponds to typical core pressure of the neutron stars, and thus \eq{EoSmain} yields sensible neutron star configurations when using it to solve the following Tolman-Oppenheimer-Volkoff (TOV) equations,
\ba
&&
\frac{dp}{dr}=-G_N(\e+p/c^2)\frac{m+4\pi r^3p/c^2}{r(r-2G_Nm/c^2)}\,,\nn\\
&&
\frac{dm}{dr}=4\pi r^2 \e \,.
\ea

   Now it is more convenient to rescale the EoS in terms of the astrophysical units listed below:    
\be
r_{\odot}=G_N M_{\odot}/c^2,\quad \epsilon_{\odot}=M_{\odot}/r_{\odot}^3, \quad p_{\odot}= c^2 \epsilon_{\odot},
\ee
for the radius of the Sun, the corresponding energy density and the pressure, respectively.  In terms of this unit system and taking the value of $\cal N$ given by \eq{calN}, we can rewrite \eq{nEoS} into the following:
\be
p/p_{\odot}={\cal A} \; p_{QCD}, \qquad   \epsilon/\epsilon_{\odot}={\cal A} \; \epsilon_{QCD}
\ee 
and
\be\label{nbvsns}
 n_b/n_s=0.12  \ell^{-5} n_I, \qquad  n_s=0.16/\textrm{fm}^3
 \ee
with 
\be\label{A&L}
{\cal A}=1.8 \times 10^{-5} \times \ell^{-7}
\ee
where $n_s$ is the the saturation density of nuclei. Then, the dimensionless EoS \eq{EoSmain} can be turned into the following dimensionful one in the astrophysical units (for T=0 case):
\be\label{EoSTOV}
\epsilon/\epsilon_{\odot}=2.629 {\cal A}^{-0.192} (p/p_{\odot})^{1.192}+0.131 {\cal A}^{0.544} (p/p_{\odot})^{0.456}.
\ee     

   We now argue that \eq{EoSTOV} yields sensible compact star configurations.  That is, we should show the maximal value of $p_{QCD}$ should be about $0.05$ for which the numerical EoS is well-fitted by the above doubly polytropic function. At the same time, we will determine the order of magnitude for the parameter $\cal A$ or $\ell^{-7}$.

   Firstly, note that the core pressure $p/p_{\odot}$ for a typical neutron star is about $10^{-3}$ to $10^{-2}$ \cite{Lattimer:2012nd}, thus ${\cal A} \; p_{QCD} \sim 10^{-3}$ to $10^{-2}$. If we set $p_{QCD} \sim {\cal O}(10^{-2})$, it implies ${\cal A}\sim  {\cal O}(10^{-1})$ or $\ell^{-7} \sim  {\cal O}(10^{4})$.  On the other hand, the typical value of $n_b$ inside a neutron star is about $2n_s$ to $10n_s$ \cite{Lattimer:2012nd}. From \eq{nbvsns} this then implies $\ell^{-5} n_I \sim {\cal O}(10)$. To further determine $\ell^{-5}$ we need the relation between $n_I$ and $p_{QCD}$ which can be obtained from the thermodynamic relation \eq{thermo} and is found to be well-fitted by the following doubly polytropic function (for $T=0$) for the same regime of $p_{QCD}\in [0, 0.05]$ in fitting \eq{EoSmain},
\be\label{nIfit}
n_I=0.617 \; p_{QCD}^{0.441}+2.300 \; p_{QCD}^{1.074}\;.
\ee
If we set  $p_{QCD} \sim {\cal O}(10^{-2})$ in \eq{nIfit}, then we get $n_I \sim {\cal O}(10^{-2})$, which then implies $\ell^{-5} \sim {\cal O}(10^3)$ or $\ell^{-7} \sim {\cal O}(10^4)$.

   From the above discussions, we can conclude that the parameter ${\cal A}\sim  {\cal O}(10^{-1})$ or equivalently $\ell^{-7} \sim  {\cal O}(10^{4})$ yields the typical values of baryon density and core pressure inside the neutron stars. Moreover, this corresponds to $p_{QCD} \sim {\cal O}(10^{-2})$ so that \eq{EoSTOV} will be the EoS for the holographic nuclear matter to yield compact neutron stars. This will be justified further by the TOV solutions obtained in the next section.  
 
\begin{figure}[htbp] 
\centering
\includegraphics[scale=0.7]{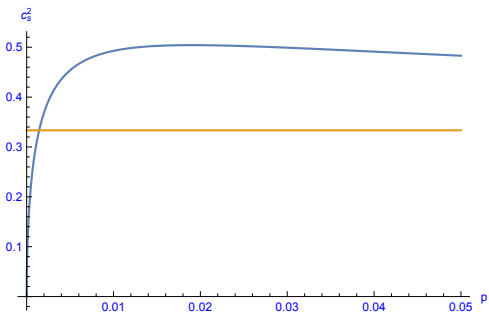}
\caption{Sound-speed-squared ($c_s^2$) vs the Pressure ($p_{QCD}$). The horizontal line is the sound barrier, i.e., $c^2_s=1/3$.}
\label{soundspeed}
\end{figure}

 Moreover, the sound speed squared $c_s^2:=\partial p/\partial \epsilon$ derived from our EoS \eq{EoSTOV} satisfies the causality constraint, i.e., $c_s < c$ and also break the sound barrier for most of the regime\footnote{The conformal barrier for the sound speed in SS model is $c_s=2c/\sqrt{5}$ for the phase of chiral restoration \cite{BitaghsirFadafan:2018uzs}, and is $c_s=c/\sqrt{3}$ for the D3/D7 model \cite{Hoyos:2016cob}.},  see Figure \ref{soundspeed}. Thus, our EoS is stiff enough to support more massive neutron stars. This is in contrast to the holographic neutron star model based on D3/D7-branes proposed in \cite{Hoyos:2016zke,Hoyos:2016cob,Annala:2017tqz}, where they need the additional inputs outside their model to break the sound barrier. 

\section{The holographic stars}

\begin{figure}[htpb]
  \centering
  \subfigure[$M$ vs $R$]{\includegraphics[height=0.24\textwidth]{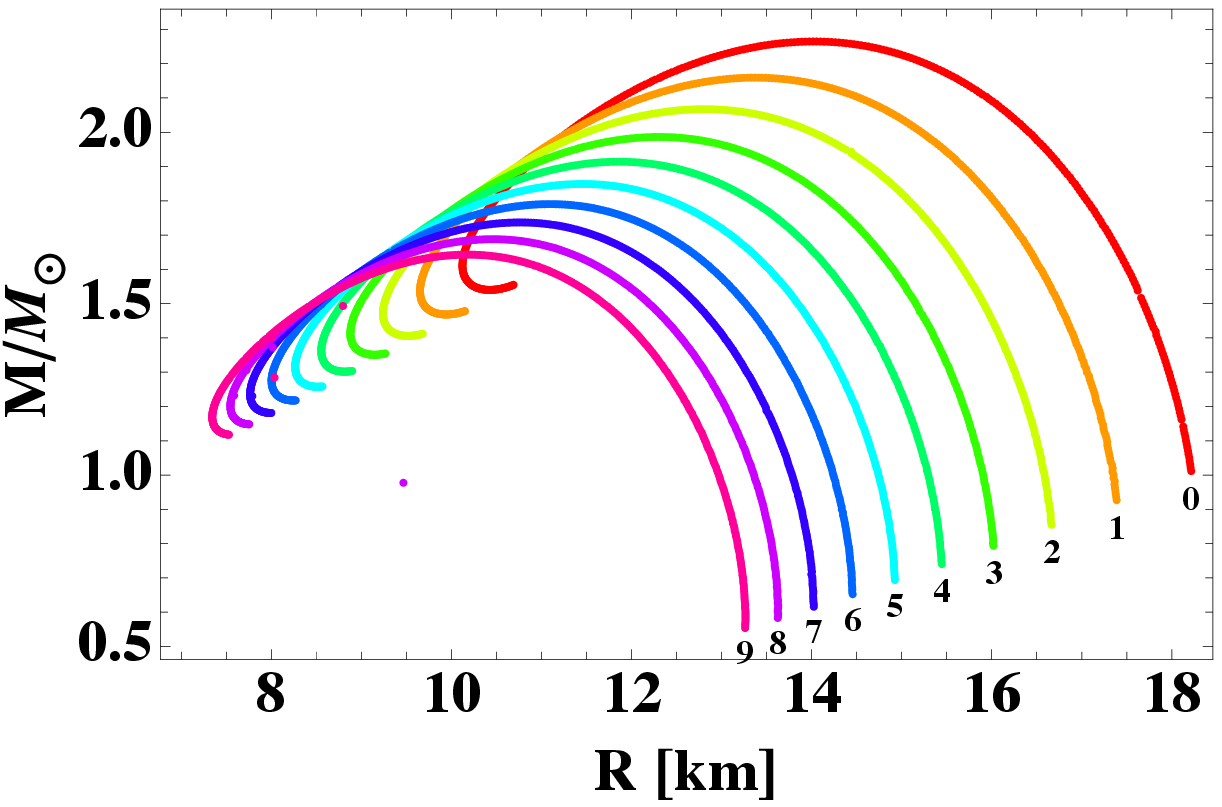}}
   \subfigure[$p_c$ vs $M$]{\includegraphics[height=0.24\textwidth]{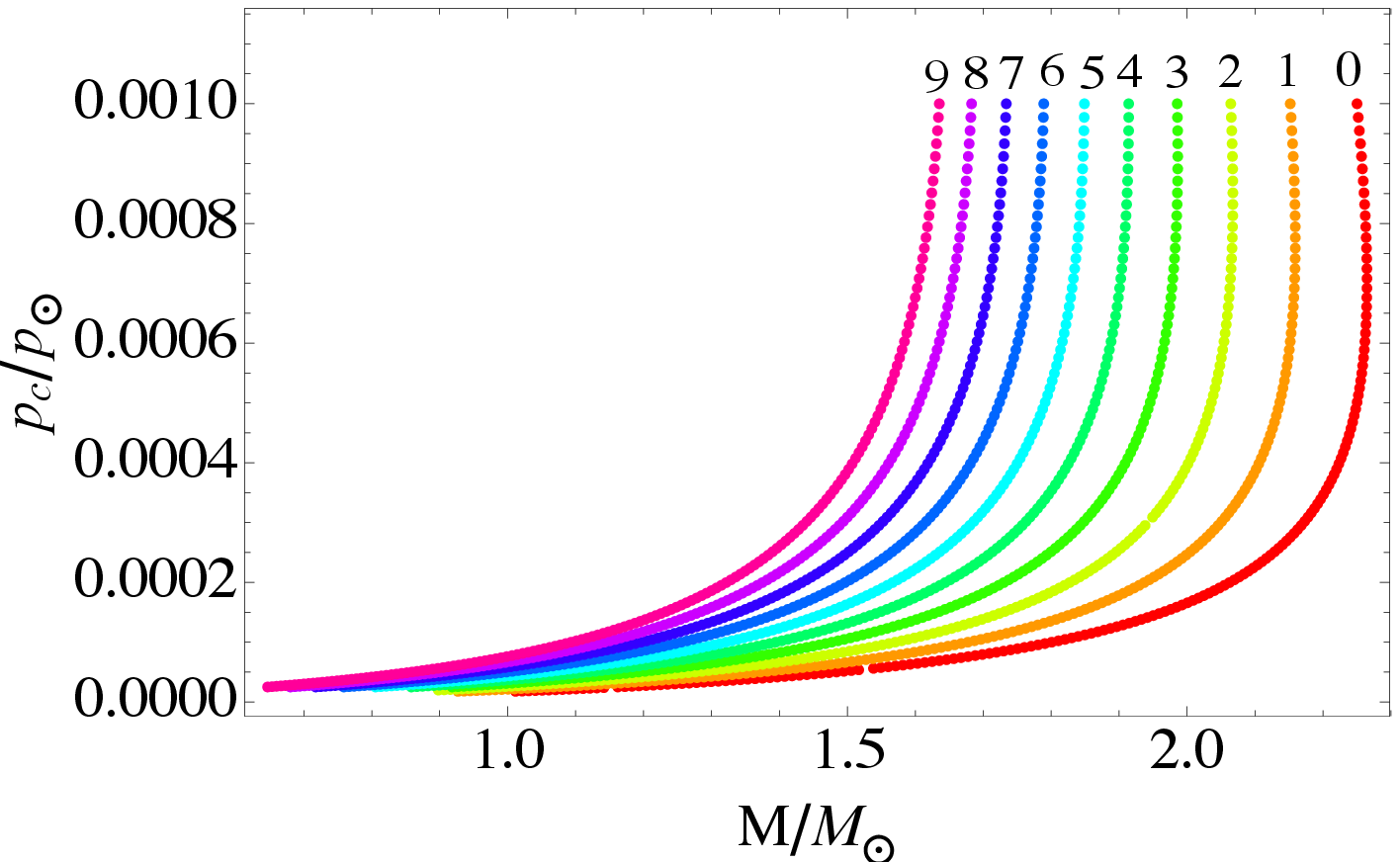}}
 \caption{(a) Mass ($M$) vs Radius ($R$)  and (b) Core-Pressure ($p_c$) vs Mass ($M$) for the holographic stars of EoS \eq{EoSTOV} with parameter $\ell^{-7}$ equally ranging from $10000$ to $19000$, which are labelled from $0$ to $9$, respectively. 
 }\label{MvR}
\end{figure}

   Based on EoS \eq{EoSTOV}, we solve TOV equations for different values of $\ell^{-7}$ with the prescribed order of magnitude around $10^4$, and then yield the mass-radius ($M$ vs $R$) relation of the holographic stars, etc. In Fig.~1 we show (a) the mass-radius relation and (b)  the relation between the core pressure and the mass ($p_c$ vs $M$), for ten values of $\ell^{-7}$ equally ranging from $10000$ to $19000$, which are labelled from $0$ to $9$, respectively. We see that the maximal mass can reach more than $2 M_{\odot}$ for $\ell^{-7} \le 13000$. In (a) of Fig.~\ref{MvR} the lowest maximal mass is about $1.62 M_{\odot}$ for $\ell^{-7}=19000$, and we expect this value will be lower if one further increases $\ell$. We choose $1.62 M_{\odot}$ because it is still larger than the upper bound $1.6 M_{\odot}$ shown in the data analysis of GW170817 \cite{TheLIGOScientific:2017qsa,Abbott:2018wiz}. Moreover, we can also infer that the compactness $M/R$ increase as $\ell^{-7}$ increases. In (b) of Fig.~\ref{MvR}  we see that the core pressure is about $10^{-3} p_{\odot}$ so that the baryon density is a few $n_s$.     
   
\begin{figure}[htpb]
  \centering
   \includegraphics[height=0.27\textwidth]{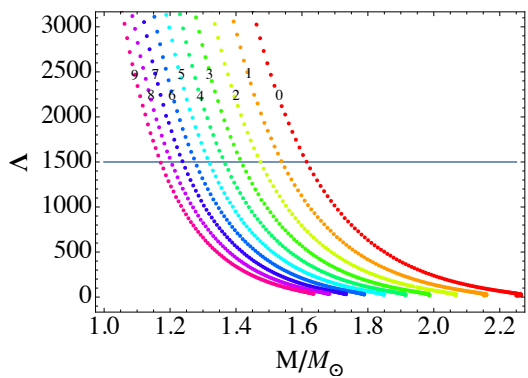}
  \caption{Tidal deformability ($\Lambda$) vs Mass ($M$) for the holographic stars of EoS \eq{EoSTOV} with the same set of values and labels for $\ell$ as in Fig.~\ref{MvR}. The middle straight line roughly indicates the upper bound on  $\Lambda$ from GW170817. The results show that it is easier to deform the less compact stars.}\label{LambdavM}
\end{figure}

    We see that our EoS satisfies the causality constraint, breaks the sound barrier, and  can be stiff enough by tuning $\ell$ to support the star with mass in excess of $2 M_{\odot}$ in some astrophysical observations \cite{Lattimer:2012nd}. We can constrain $\ell$ further by also evaluating the tidal deformability and compare with the inference values from observation data of GW170817 \cite{TheLIGOScientific:2017qsa,Abbott:2018wiz}.   The tidal deformability characterizes how the shape of the star is deformed by the external gravitational field, and is defined as the dimensionless coefficient $\Lambda$ in the following linear response relation
\be
Q_{ij}=- \left({M\over M_{\odot}}\right)^5  \Lambda \; {\cal E}_{ij}
\ee
where $M$ is the mass of the star, $Q_{ij}$ is the induced quadrupole moment, and ${\cal E}_{ij}$ is the external gravitational tidal field strength.    Given the EoS and a neutron star configuration, we can follow the perturbative method of \cite{Hinderer:2007mb} to calculate the tidal deformability. For our EoS \eq{EoSTOV} and the star configurations shown in Fig.~\ref{MvR}, the relation of tidal deformability and mass ($\Lambda$ vs $M$) is shown in Fig.~\ref{LambdavM}. We see that $\Lambda$ increases as $\ell^{-7}$ decreases, this implies that it is easier to deform for less compact star as  intuitively expected. 
 
\begin{figure}[htpb]
  \centering
   \includegraphics[height=0.25\textwidth]{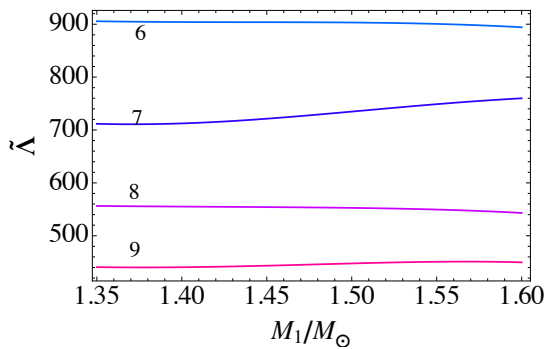}
  \caption{Tidal deformability of the binary holographic stars of EoS \eq{EoSTOV} vs one of the masses ($\tilde{\lambda}$ vs $M_1$) for partial set of the values for $\ell$ used in Fig.~\ref{MvR}. The ones labelled by $7$, $8$ and $9$ are consistent with the estimated value $300^{+420}_{-230}$ of GW170817.}\label{tLambdavM1}
\end{figure}

    We see from Fig.~\ref{LambdavM} that $\Lambda$ can cover a large range. However, the data analysis of GW170817 \cite{TheLIGOScientific:2017qsa,Abbott:2018wiz} shows that the tidal deformability is moderately constrained. As GW170817 is a system of binary neutron stars, the data is fitted for the following combined quantity
\be\label{tlambda}
\tilde{\Lambda}={16\over 13}{(M_1+12 M_2) M_1^4 \Lambda_1+(M_2+12 M_1) M_2^4 \Lambda_2 \over (M_1+M_2)^5}
\ee 
where $M_{1,2}$ are the masses of two neutron stars with $M_1<M_2$, and $\Lambda_{1,2}$ are their associated tidal deformabilities.   The analysis of GW170817 in  \cite{TheLIGOScientific:2017qsa,Abbott:2018wiz}  yields an estimate on $\tilde{\Lambda}=300^{+420}_{-230}$, which implies the upper bound of $\Lambda$ is about 1500, according to \eqref{tlambda}.  This will serve as a further constraint on $\ell^{-7}$ of our EoS.

  To obtain $\tilde{\Lambda}$ for our EoS and star configurations, we first fit the curve of $M_1$ vs $M_2$ for the low-spin prior in Fig. 5 of \cite{Abbott:2018wiz} with $M_1+M_2\approx 2.7 M_{\odot}$, and then plug into \eq{tlambda} to get $\tilde{\Lambda}$ vs $M_1$. The result is shown in Fig.~\ref{tLambdavM1}. We see that the $\Lambda$'s in Fig.~\ref{LambdavM} are too large for lower $\ell^{-7}$ so that we present only the ones closer to the estimate of \cite{TheLIGOScientific:2017qsa,Abbott:2018wiz}. From Fig.~\ref{tLambdavM1} we see that only $\ell^{-7}=17000$, $18000$ and  $19000$ are consistent with the observation. For these cases, the predicted radii shown in Fig.~\ref{MvR} are consistent with the inferred values of LIGO/Virgo observation of GW170817, i.e., $\sim 11  \textrm{km}$ \cite{Abbott:2018exr}.  However, the maximal mass is about $1.7 M_{\odot}$, which encompasses the mass estimates of most observed NSs in various types of NS-containing systems (e.g., double-NS systems, NS-white dwarf binaries, X-ray binaries) \cite{Lattimer:2012nd},  but is still smaller than some observed values, especially the ones in excess of $2 M_{\odot}$. On the other hand, we can use the upper bound of the tidal deformability, say $\Lambda\sim 1500$ to obtain the lowest mass from Fig.~\ref{LambdavM}, and it is about $1.2 M_{\odot}$ for $\ell^{-7}=17000$ to $19000$.  This is consistent with the observed values \cite{Fonseca:2016tux,Martinez:2015mya,Falanga:2015mra}.

 \section{Conclusion}   
 
   In this work we give a first principle derivation of nuclear matter EoS based on a top-down model of holographic QCD. The resultant EoS is doubly polytropic with only one free parameter, and can yield mass, radius and tidal deformability in excellent agreement with observation of GW170817.  However, our EoS cannot be consistent with GW170817 and yield the maximal mass in excess of $2 M_{\odot}$ at the same time.  This leaves the space for future studies by considering more general instanton profiles other than a delta function \cite{Li:2015uea,BitaghsirFadafan:2018uzs}, and  considering twin stars \cite{Oestgaard:1994gy} with hybrid EoS of holographic baryonic and quark matters. The future events of NS binaries observed in LIGO/Virgo/KAGRA should help to pin down the necessity of the above options to reach the higher maximal mass of NSs.

~

\noindent\textit{%
FLL is supported by Taiwan Ministry of Science and Technology (MoST) through Grant No.~103-2112-M-003-001-MY3. LWL is supported by Academia Sinica Career Development Award Program through Grant No.~AS-CDA-105-M06. KZ(Hong Zhang) thanks Yutaka Matsuo for useful advice and is supported by MoST through Grant No.~107-2811-M-003-511. We thank Alessandro Parisi, Meng-Ru Wu for helpful discussions. We also thank NCTS for partial financial support. 
}


\begin{thebibliography}{0}%
\makeatletter
\providecommand \@ifxundefined [1]{%
 \@ifx{#1\undefined}
}%
\providecommand \@ifnum [1]{%
 \ifnum #1\expandafter \@firstoftwo
 \else \expandafter \@secondoftwo
 \fi
}%
\providecommand \@ifx [1]{%
 \ifx #1\expandafter \@firstoftwo
 \else \expandafter \@secondoftwo
 \fi
}%
\providecommand \natexlab [1]{#1}%
\providecommand \enquote  [1]{``#1''}%
\providecommand \bibnamefont  [1]{#1}%
\providecommand \bibfnamefont [1]{#1}%
\providecommand \citenamefont [1]{#1}%
\providecommand \href@noop [0]{\@secondoftwo}%
\providecommand \href [0]{\begingroup \@sanitize@url \@href}%
\providecommand \@href[1]{\@@startlink{#1}\@@href}%
\providecommand \@@href[1]{\endgroup#1\@@endlink}%
\providecommand \@sanitize@url [0]{\catcode `\\12\catcode `\$12\catcode
  `\&12\catcode `\#12\catcode `\^12\catcode `\_12\catcode `\%12\relax}%
\providecommand \@@startlink[1]{}%
\providecommand \@@endlink[0]{}%
\providecommand \url  [0]{\begingroup\@sanitize@url \@url }%
\providecommand \@url [1]{\endgroup\@href {#1}{\urlprefix }}%
\providecommand \urlprefix  [0]{URL }%
\providecommand \Eprint [0]{\href }%
\providecommand \doibase [0]{http://dx.doi.org/}%
\providecommand \selectlanguage [0]{\@gobble}%
\providecommand \bibinfo  [0]{\@secondoftwo}%
\providecommand \bibfield  [0]{\@secondoftwo}%
\providecommand \translation [1]{[#1]}%
\providecommand \BibitemOpen [0]{}%
\providecommand \bibitemStop [0]{}%
\providecommand \bibitemNoStop [0]{.\EOS\space}%
\providecommand \EOS [0]{\spacefactor3000\relax}%
\providecommand \BibitemShut  [1]{\csname bibitem#1\endcsname}%
\let\auto@bib@innerbib\@empty
\end{thebibliography}%


\begin{thebibliography}{99}%

 
\bibitem{Lattimer:2012nd}
  J.~M.~Lattimer,
  ``The nuclear equation of state and neutron star masses,''
  Ann.\ Rev.\ Nucl.\ Part.\ Sci.\  {\bf 62} (2012) 485
  [arXiv:1305.3510 [nucl-th]].



 
\bibitem{Baym:2019yyo} 
  G.~Baym,
  ``The Golden Era of Neutron Stars: from Hadrons to Quarks,''
  arXiv:1902.01274 [astro-ph.HE].

\bibitem{Rezzolla:2016nxn} 
  L.~Rezzolla and K.~Takami,
  ``Gravitational-wave signal from binary neutron stars: a systematic analysis of the spectral properties,''
  Phys.\ Rev.\ D {\bf 93}, no. 12, 124051 (2016)
  [arXiv:1604.00246 [gr-qc]].


\bibitem{TheLIGOScientific:2017qsa} 
  B.~P.~Abbott {\it et al.} [LIGO Scientific and Virgo Collaborations],
  ``GW170817: Observation of Gravitational Waves from a Binary Neutron Star Inspiral,''
  Phys.\ Rev.\ Lett.\  {\bf 119}, no. 16, 161101 (2017)
  [arXiv:1710.05832 [gr-qc]].

\bibitem{Most:2018hfd} 
  E.~R.~Most, L.~R.~Weih, L.~Rezzolla and J.~Schaffner-Bielich,
  ``New constraints on radii and tidal deformabilities of neutron stars from GW170817,''
  Phys.\ Rev.\ Lett.\  {\bf 120}, no. 26, 261103 (2018)
  [arXiv:1803.00549 [gr-qc]].


\bibitem{Abbott:2018wiz} 
  B.~P.~Abbott {\it et al.} [LIGO Scientific and Virgo Collaborations],
  ``Properties of the binary neutron star merger GW170817,''
  Phys.\ Rev.\ X {\bf 9}, no. 1, 011001 (2019)
  [arXiv:1805.11579 [gr-qc]].



\bibitem{Abbott:2018exr} 
  B.~P.~Abbott {\it et al.} [LIGO Scientific and Virgo Collaborations],
  ``GW170817: Measurements of neutron star radii and equation of state,''
  Phys.\ Rev.\ Lett.\  {\bf 121}, no. 16, 161101 (2018)
  [arXiv:1805.11581 [gr-qc]].
 
 
 
 
  
 
\bibitem{Tews:2018chv} 
  I.~Tews, J.~Margueron and S.~Reddy,
  ``Critical examination of constraints on the equation of state of dense matter obtained from GW170817,''
  Phys.\ Rev.\ C {\bf 98}, no. 4, 045804 (2018)
  [arXiv:1804.02783 [nucl-th]].
 
 
\bibitem{De:2018uhw} 
  S.~De, D.~Finstad, J.~M.~Lattimer, D.~A.~Brown, E.~Berger and C.~M.~Biwer,
  ``Tidal Deformabilities and Radii of Neutron Stars from the Observation of GW170817,''
  Phys.\ Rev.\ Lett.\  {\bf 121}, no. 9, 091102 (2018)
  Erratum: [Phys.\ Rev.\ Lett.\  {\bf 121}, no. 25, 259902 (2018)]
  [arXiv:1804.08583 [astro-ph.HE]].
 
 
 
   
\bibitem{Zhao:2018nyf} 
  T.~Zhao and J.~M.~Lattimer,
  ``Tidal Deformabilities and Neutron Star Mergers,''
  Phys.\ Rev.\ D {\bf 98}, no. 6, 063020 (2018)
  [arXiv:1808.02858 [astro-ph.HE]].
  

   
\bibitem{Han:2018mtj} 
  S.~Han and A.~W.~Steiner,
  ``Tidal deformability with sharp phase transitions in (binary) neutron stars,''
  arXiv:1810.10967 [nucl-th].
  
  
 
\bibitem{Carson:2018xri} 
  Z.~Carson, A.~W.~Steiner and K.~Yagi,
  ``Constraining nuclear matter parameters with GW170817,''
  arXiv:1812.08910 [gr-qc].
  
  

  
\bibitem{Tews:2019cap} 
  I.~Tews, J.~Margueron and S.~Reddy,
  ``Confronting gravitational-wave observations with modern nuclear physics constraints,''
  arXiv:1901.09874 [nucl-th].
  
 
   
 
 
\bibitem{deForcrand:2010ys} 
  P.~de Forcrand,
  ``Simulating QCD at finite density,''
  PoS LAT {\bf 2009}, 010 (2009)
  [arXiv:1005.0539 [hep-lat]].
 
\bibitem{Cristoforetti:2012su} 
  M.~Cristoforetti, F.~DiRenzo, L.~Scorzato,
  ``New approach to the sign problem in quantum field theories: High density QCD on a Lefschetz thimble,''
  Phys.\ Rev.\ D {\bf 86}, 074506 (2012)
  [arXiv:1205.3996 [hep-lat]].
 
 
 
 
\bibitem{Kraemmer:2003gd} 
  U.~Kraemmer and A.~Rebhan,
  ``Advances in perturbative thermal field theory,''
  Rept.\ Prog.\ Phys.\  {\bf 67}, 351 (2004)
  [hep-ph/0310337].
 
\bibitem{Kurkela:2016was} 
  A.~Kurkela and A.~Vuorinen,
  ``Cool quark matter,''
  Phys.\ Rev.\ Lett.\  {\bf 117}, no. 4, 042501 (2016)
  [arXiv:1603.00750 [hep-ph]].
  
\bibitem{Vuorinen:2018qzx} 
  A.~Vuorinen,
  ``Neutron stars and stellar mergers as a laboratory for dense QCD matter,''
  Nucl.\ Phys.\ A {\bf 982}, 36 (2019)
  [arXiv:1807.04480 [nucl-th]].

   
\bibitem{Maldacena:1997re} 
  J.~M.~Maldacena,
  ``The Large N limit of superconformal field theories and supergravity,''
  Int.\ J.\ Theor.\ Phys.\  {\bf 38}, 1113 (1999)
  [Adv.\ Theor.\ Math.\ Phys.\  {\bf 2}, 231 (1998)]
  [hep-th/9711200].
   
   
\bibitem{Kovtun:2004de} 
  P.~K.~Kovtun, D.~T.~Son and A.~O.~Starinets,
  ``Viscosity in strongly interacting quantum field theories from black hole physics,''
  Phys.\ Rev.\ Lett.\  {\bf 94}, 111601 (2005)
  [hep-th/0405231].
   
\bibitem{Liu:2006ug} 
  H.~Liu, K.~Rajagopal and U.~A.~Wiedemann,
  Phys.\ Rev.\ Lett.\  {\bf 97}, 182301 (2006)
  doi:10.1103/PhysRevLett.97.182301
  [hep-ph/0605178].
  
   
\bibitem{Gubser:2006bz} 
  S.~S.~Gubser,
  ``Drag force in AdS/CFT,''
  Phys.\ Rev.\ D {\bf 74}, 126005 (2006)
  [hep-th/0605182].

   
\bibitem{Gynther:2010ed} 
  A.~Gynther, K.~Landsteiner, F.~Pena-Benitez and A.~Rebhan,
  ``Holographic Anomalous Conductivities and the Chiral Magnetic Effect,''
  JHEP {\bf 1102}, 110 (2011)
  [arXiv:1005.2587 [hep-th]].

   
   
\bibitem{Sakai:2004cn} 
  T.~Sakai and S.~Sugimoto,
  ``Low energy hadron physics in holographic QCD,''
  Prog.\ Theor.\ Phys.\  {\bf 113}, 843 (2005)
  doi:10.1143/PTP.113.843
  [hep-th/0412141].
  
  
  

\bibitem{Sakai:2005yt} 
  T.~Sakai and S.~Sugimoto,
  ``More on a holographic dual of QCD,''
  Prog.\ Theor.\ Phys.\  {\bf 114}, 1083 (2005)
  doi:10.1143/PTP.114.1083
  [hep-th/0507073].
  
  
  
  
\bibitem{Witten:1998zw} 
  E.~Witten,
  ``Anti-de Sitter space, thermal phase transition, and confinement in gauge theories,''
  Adv.\ Theor.\ Math.\ Phys.\  {\bf 2}, 505 (1998)
  doi:10.4310/ATMP.1998.v2.n3.a3
  [hep-th/9803131].


\bibitem{Hata:2007mb} 
  H.~Hata, T.~Sakai, S.~Sugimoto and S.~Yamato,
  ``Baryons from instantons in holographic QCD,''
  Prog.\ Theor.\ Phys.\  {\bf 117}, 1157 (2007)
  doi:10.1143/PTP.117.1157
  [hep-th/0701280 [HEP-TH]].




     
\bibitem{Witten:1998xy} 
  E.~Witten,
  ``Baryons and branes in anti-de Sitter space,''
  JHEP {\bf 9807}, 006 (1998)
  doi:10.1088/1126-6708/1998/07/006
  [hep-th/9805112].
  
  
\bibitem{Bergman:2007wp} 
  O.~Bergman, G.~Lifschytz and M.~Lippert,
  ``Holographic Nuclear Physics,''
  JHEP {\bf 0711}, 056 (2007)
  doi:10.1088/1126-6708/2007/11/056
  [arXiv:0708.0326 [hep-th]].

 
  
  
  
  
\bibitem{Antonyan:2006vw} 
  E.~Antonyan, J.~A.~Harvey, S.~Jensen and D.~Kutasov,
  ``NJL and QCD from string theory,''
  hep-th/0604017.
  
  
\bibitem{Davis:2007ka} 
  J.~L.~Davis, M.~Gutperle, P.~Kraus and I.~Sachs,
  ``Stringy NJL and Gross-Neveu models at finite density and temperature,''
  JHEP {\bf 0710}, 049 (2007)
  doi:10.1088/1126-6708/2007/10/049
  [arXiv:0708.0589 [hep-th]].
  
      
    
\bibitem{Li:2015uea} 
  S.~W.~Li, A.~Schmitt and Q.~Wang,
  ``From holography towards real-world nuclear matter,''
  Phys.\ Rev.\ D {\bf 92}, no. 2, 026006 (2015)
  [arXiv:1505.04886 [hep-ph]].
 
\bibitem{BitaghsirFadafan:2018uzs} 
  K.~Bitaghsir Fadafan, F.~Kazemian and A.~Schmitt,
  ``Towards a holographic quark-hadron continuity,''
  arXiv:1811.08698 [hep-ph].
  
 
 
 
  


 
  
\bibitem{Hoyos:2016zke} 
  C.~Hoyos, N.~Jokela, D.~Rodriguez Fernandez and A.~Vuorinen,
  ``Holographic quark matter and neutron stars,''
  Phys.\ Rev.\ Lett.\  {\bf 117}, no. 3, 032501 (2016)
  [arXiv:1603.02943 [hep-ph]].
 
 
\bibitem{Hoyos:2016cob} 
  C.~Hoyos, N.~Jokela, D.~Rodriguez Fernandez and A.~Vuorinen,
  ``Breaking the sound barrier in AdS/CFT,''
  Phys.\ Rev.\ D {\bf 94}, no. 10, 106008 (2016)
  [arXiv:1609.03480 [hep-th]].
 
\bibitem{Annala:2017tqz} 
  E.~Annala, C.~Ecker, C.~Hoyos, N.~Jokela, D.~Rodríguez Fernández and A.~Vuorinen,
  ``Holographic compact stars meet gravitational wave constraints,''
  JHEP {\bf 1812}, 078 (2018)
  [arXiv:1711.06244 [astro-ph.HE]].
 
 
 
\bibitem{Hinderer:2007mb} 
  T.~Hinderer,
  ``Tidal Love numbers of neutron stars,''
  Astrophys.\ J.\  {\bf 677}, 1216 (2008)
  [arXiv:0711.2420 [astro-ph]].
 
 
\bibitem{Fonseca:2016tux} 
  E.~Fonseca {\it et al.},
  ``The NANOGrav Nine-year Data Set: Mass and Geometric Measurements of Binary Millisecond Pulsars,''
  Astrophys.\ J.\  {\bf 832}, no. 2, 167 (2016)
  [arXiv:1603.00545 [astro-ph.HE]].
 
\bibitem{Martinez:2015mya} 
  J.~G.~Martinez {\it et al.},
  ``Pulsar J0453+1559: A Double Neutron Star System with a Large Mass Asymmetry,''
  Astrophys.\ J.\  {\bf 812}, no. 2, 143 (2015)
  [arXiv:1509.08805 [astro-ph.HE]].
 
\bibitem{Falanga:2015mra} 
  M.~Falanga, E.~Bozzo, A.~Lutovinov, J.~M.~Bonnet-Bidaud, Y.~Fetisova and J.~Puls,
  ``Ephemeris, orbital decay, and masses of ten eclipsing high-mass X-ray binaries,''
  Astron.\ Astrophys.\  {\bf 577}, A130 (2015)
  [arXiv:1502.07126 [astro-ph.HE]].
 

\bibitem{Oestgaard:1994gy} 
  E.~Oestgaard,
  ``Compact stars: Neutron stars or quark stars or hybrid stars?,''
  Phys.\ Rept.\  {\bf 242}, 313 (1994).




 
 
 \end{thebibliography}
 \end{document}